\documentclass[prd,twocolumn,showpacs,floats,letterpaper,nofootinbib]{revtex4}

\usepackage{amssymb,amsmath,latexsym,mathrsfs}
\usepackage{graphicx,subfigure}
\usepackage{epsfig}
\usepackage{varioref,xr-hyper}
\usepackage{color}

\usepackage[hypertex,draft=false,verbose=false,debug=false
            hyperindex=true,hypertexnames=true]{hyperref}

\newcommand{\hide}[1]{{}}

\newcommand{\be}{\begin{equation}}
\newcommand{\ee}{\end{equation}}
\newcommand{\bea}{\begin{eqnarray}}
\newcommand{\eea}{\end{eqnarray}}

\newcommand{\begm}{\begin{pmatrix}}
\newcommand{\enm}{\end{pmatrix}}

\def\lsim{\;\raise 0.4ex\hbox{$<$}\kern -0.8em\lower 0.62 ex\hbox{$\sim$}\;}
\def\gsim{\;\raise 0.4ex\hbox{$>$}\kern -0.7em\lower 0.62 ex\hbox{$\sim$}\;}

\begin{document}

\title{CMB Lensing Constraints on Dark Energy and Modified Gravity Scenarios}

\author{Erminia Calabrese$^{1,2}$, Asantha Cooray$^{2}$, Matteo Martinelli$^{1,3}$,\\
Alessandro Melchiorri$^{1}$, Luca Pagano$^{1,4}$,
An\v{z}e Slosar$^{5}$, George F. Smoot$^{5,6,7}$}

\

\affiliation{$^1$Physics Department and INFN, Universita' di Roma ``La Sapienza'',
  Ple Aldo Moro 2, 00185, Rome, Italy}
\affiliation{$^2$Center for Cosmology, Dept. of Physics \& Astronomy, University of California Irvine, Irvine, CA 92697.}
\affiliation{$^3$Department of Astrophysical Sciences, Princeton University, Princeton, NJ 08544, USA.}
\affiliation{$^4$Jet Propulsion Laboratory, California Institute of Technology, 4800 Oak Grove Drive, Pasadena CA 91109, USA.}
\affiliation{$^5$Berkeley Center for Cosmological Physics, Lawrence Berkeley National Laboratory and Physics Department, University of California, Berkeley CA 94720.}
\affiliation{$^6$Institute for the Early Universe, Ewha Womans University and Ewha Advanced Academy, Seoul, Korea.}
\affiliation{$^7$Chaire Blaise Pascal, Universite' Paris Diderot - 75205 PARIS cedex 13.}


\begin{abstract}
Weak gravitational lensing leaves a characteristic imprint on the
cosmic microwave background temperature and polarization angular power spectra.
Here we investigate the possible constraints on the integrated lensing potential
 from future CMB angular spectra measurements expected from Planck and EPIC.
 We find that Planck and EPIC will constrain the amplitude
 of the integrated projected potential responsible for lensing at $6 \%$ and $1\%$ level, respectively
 with very little sensitivity to the shape of the lensing potential.
 We discuss the implications of such a measurement in constraining dark energy
 and modified gravity scalar-tensor theories.  We then discuss the impact of a wrong assumption on the weak lensing potential
amplitude on cosmological parameter inference.
\end{abstract}


\maketitle

\section{Introduction}

Due to extremely sensitive observations with
 satellite (\cite{wmap5}), ground-based (\cite{cbi}, \cite{acbar}) and balloon-borne experiments (\cite{boom03}),
the Cosmic Microwave Background (CMB hereafter) temperature and polarization anisotropies are now measured with astonishing
precision.  Moreover, these measurements are in nearly perfect agreement with theoretical predictions of an adiabatic
cold dark matter (CDM) model of structure formation and most of the CDM cosmological parameters are
now constrained at better than a few percent uncertainty.

Future data expected from Planck (\cite{planck}) and ground-based
experiments such as SPT (\cite{spt}) will provide even better measurements with cosmic variance limited maps
of CMB temperature anisotropies down to few arcminutes angular resolution.
Moreover, balloon-borne experiments such as SPIDER (\cite{spider}) and EBEX (\cite{ebex}),
and possible future satellite mission EPIC (\cite{epic}) will provide further constraints on
the polarization signal, again at the cosmic variance limit of polarization measurements down to ten arcminute angular scale or better.

With all this high quality data expected in the very near future it is
definitely timely to investigate if the theoretical predictions have reached a similar
accuracy.

The calculation of CMB anisotropies through current Boltzmann solvers as CMBFAST \cite{cmbfast}
or CAMB \cite{camb} is made in the linear regime, where the evolution equations can in principle be
solved to an arbitrary precision. Current codes show deviations at $\sim 0.1 \%$ level, well
below the future experimental accuracy (see e.g. \cite{accuracy}).

However, additional uncertainties are present. The thermal history of primordial
recombination, for example, is still not sufficiently theoretically
determined to match future observations (see e.g. \cite{hirata}).

While the needed accuracy in this case relies on standard physics and it is expected to be achievable,
a more worrying aspect concerns the assumptions on the unknown dark energy component.
The choice of the dark energy framework, indeed, clearly modifies the CMB anisotropy angular spectrum. It is well known that
the dark energy density and its equation of state $w$ defined as the ratio between pressure
and energy of the dark energy fluid both change the angular diameter distance of the sound
horizon at last scattering and shift the peaks positions in the CMB spectra (see e.g. \cite{bean}).
These parameters can be accurately measured by the
use of several complementary observables such as luminosity distances of type Ia supernovae (\cite{kowalski}).

It has however recently become clear that models of dark energy
or modified gravity can be conceived leading to the same background expansion of $\Lambda-CDM$ (and therefore
preserving the angular diameter distance at redshift $z\sim 1100$) but with different evolution for the
perturbations in the energy components (see e.g  \cite{silvestri}, \cite{marcald} and references therein).

The CMB angular spectra in those models may have small but measurable differences respect to those expected in the $\Lambda-CDM$ model.
First, on large angular scales, since the growth of perturbations is different, gravitational potentials will
change differently with time, affecting the CMB through the so-called Integrated Sachs-Wolfe (ISW) effect \cite{isw}.
Recent works (see e.g. \cite{song07}, \cite{caldwell}) have focused on this large angular scale anisotropy signal
in order to extract information on dark energy or modified gravity. It is clear that one could invert the argument
and ask to what extent our poor theoretical knowledge of the ISW signal could affect cosmic parameters. Fortunately this
signal is dominated by the primordial anisotropy and present only on the very few first multipoles.
In few words, the impact on parameter extraction of a wrong theoretical prediction for the ISW signal is
negligible.

However, dark energy also changes the small angular scale anisotropy through lensing.
While the physics of CMB lensing is well understood, the amplitude of the signal is indeed connected to
the growth of matter density fluctuations with redshift. Since the growth of structure
strongly depends on the dark energy component and since a clear physical understanding of this component is lacking,
it is definitely possible that the lensing amplitude will be different from the one expected in the
cosmological constant scenario.

Several recent papers have shown that the lensed CMB signal may be used to
study dark energy \cite{Hu01c,Kap03,AcqBac05,serra09}. Here we follow
a complementary approach, discussing the impact of an unaccounted variation in the lensing signal,
motivated either by a different dark energy model or modified gravity, on cosmological parameter inference.
As we show in the next section, modified gravity, for example, could easily
increase the rms lensing signal by $\sim 20 \%$ while leaving the primary anisotropies unaffected.
Smaller but still sizable variations in the amplitude could be induced by perturbations in the dark energy fluid.
We will show that if future data will be analyzed without considering the possibility of
a non-standard lensing signal, this may drastically bias the conclusions on many cosmological parameters.

Our paper is organized as follows: in the next Section we show
how dark energy perturbations and/or modified gravity could change the amplitude
of the weak lensing signal.
In Section III we present our data analysis method and the datasets considered. In Section IV we present
the results and in Section V we conclude with our conclusions.

\section{The amplitude of the weak lensing potential in non standard models of structure formation}

The lensing deflection of CMB photons depends
upon gradients in the total gravitational potential $\Phi+\Psi$ transverse to the
line of sight to the last scattering surface \cite{Lewis:2006fu}, where
 $\Phi$ and $\Psi$ are defined by the perturbed Robertson-Walker line-element
\begin{equation}
\label{metric}
ds^2=a^2 \left[-\left(1+2\Psi\right)d\tau^2
+\left(1-2\Phi\right)d\vec{x}^2\right] \, ,
\end{equation}
using the notation and convention of Ref.~\cite{Ma:1995ey}.

The evolution of the gravitational potential can be expressed by a
a transfer function $T_{\Psi}(\vec{k},\tau)$, whereby
$\Psi(\vec{k},\tau)=T_{\Psi}(k,\tau)R(\vec{k})$ and where $R(\vec{k})$ is the
primordial curvature perturbation.
The power spectrum of the lensing potential is given by
\begin{equation}
C_\ell^{\psi}=4\pi\int\frac{dk}{k}P_R(k)
\Big[\int_0^{\chi_*}d\chi\,S_{\Psi}(k;\tau_0-\chi)j_\ell(k\chi)\Big]^2 \, .
\label{cl_lens}
\end{equation}
Here $P_R(k)$ is the primordial power spectrum, $\tau_0-\chi$ is the conformal
time at which a given photon was at the position $\chi\hat{n}$, and the lensing
source, in the standard scenario, is given by (see e.g. \cite{lensteo}):
\begin{equation}
S_{\psi}(k,\tau_0-\chi)=2T_{\Psi}(k,\tau_0-\chi)
j_\ell(k\chi)\Big(\frac{\chi_*-\chi}{\chi_*\chi}\Big) \, .
\end{equation}

The effects of lensing on temperature and polarization anisotropy
have been extensively presented in the literature.
In the case of temperature, lensing modifies the damping tail,
smearing the acoustic oscillations. A similar effect but more pronounced
is present in the EE polarization and TE cross temperature-polarization
spectra, while lensing also introduces an extra B-mode polarization
signal. The lensed B-mode peaks at tens of  arcminute angular scales and
its amplitude is directly proportional to the lensing power spectrum.
It is clear from the expression of the lensing angular spectrum in Equation \ref{cl_lens} that
the lensing amplitude will depend on the growth of dark matter perturbations.
The evolution of the dark matter perturbations can be altered by the dark energy component,
affecting the weak lensing signal.

While the equation of state of dark energy $w$ can be constrained by complementary observables,
the dark energy perturbation sound speed, $c_s^2$ (see e.g. \cite{beandore}),
that probes the nature of dark energy fluctuations, is more elusive. It is useful therefore to investigate
the impact of a different choice for $c_s^2$ on the weak lensing signal.
Fixing $w=-0.8$, that is already at the border of current measurements that are
pointing towards $w=-1\pm0.1$, a $\sim 5 \%$ difference
is present between the lensing potential with $c_s^2=1$ and $c_s^2=0$.
Perturbations in the dark energy will be present only if the dark energy equation of state will be different from
$w=-1$ and closer the equation of state will be to this value, smaller will be the effect.
One could therefore argue that if future measurements will accurately determine that the
equation of state is extremely close to $-1$ then there will be no expected
modification to the weak lensing signal\footnote{This is not completely true since if dark energy interacts with
dark matter then perturbations will be possible even if the, effective, measured
value of $w$ will be close to $-1$}.

However, if modified gravity is at works, it is possible to obtain a background expansion
that mimics exactly $\Lambda-CDM$ but still with a different lensing amplitude.
Several modified gravity models have been proposed. Here we follow the approach presented
in \cite{alessandra} where a modified gravity changes the relations between
the two Newtonian potentials and between the same potentials and matter perturbations.
Modified gravity can indeed be parameterized with two functions $\gamma$ and $\mu$ (see \cite{alessandra}) such that:
\begin{equation}
k^2\Psi=-\frac{a^2}{2M_p^2}\mu(a,k)\rho\Delta \, ,
\end{equation}
and
\begin{equation}
\frac{\Phi}{\Psi}=\gamma(a,k)  \,,
\end{equation}
where $\rho\Delta$ is the co-moving density perturbation and $M_p$ is the Planck mass.

The functions $\mu$ and $\gamma$ encode the effects due to modified gravity or dark energy and can be expressed as
\begin{equation}
\mu(a,k)=\frac{1+\beta_1\lambda_1^2a^s}{1+\lambda_1^2a^s} \, ,
\end{equation}
and
\begin{equation}
\gamma(a,k)=\frac{1+\beta_2\lambda_2^2a^s}{1+\lambda_2^2a^s} \, ,
\end{equation}
respectively, where $\lambda_1$ is the Compton wavelength of the modified force, $\beta_1$ and $\beta_2$
 are linked to the coupling of the scalar field with matter and $s$ sets the time-dependence of $\mu$ and $\gamma$.

Referring to scalar-tensor theories, these parameters satisfy the equations
\begin{equation}
\beta_1=\frac{\lambda_2^2}{\lambda_1^2}=2-\beta_2\frac{\lambda_2^2}{\lambda_1^2}.
\end{equation}

If we choose a particular value for $\beta_1$, $\beta_2$ and $s$ we specify a particular class of theories;
for example. we can consider  $\beta_1=4/3$, $\beta_2=1/2$ and $s=4$ and
work with a general $f(R)$ theory \cite{alessandra}
where $\lambda_1$ can be related to the expression of $f(R)$ thanks to the definition of the mass related to
the modified force $m^2=(1+f_R)/f_{RR}$ \cite{HS}:
\begin{equation}
\lambda_1^2=\frac{1}{m_0^2}=\left(\frac{3f_{RR}}{1+f_R}\right)_0 \, ,
\end{equation}
where $f_R=df/dR$, $f_{RR}=d^2f/dR^2$ and the subscript $0$ denotes quantities evaluated at the present time.

Using a modified version of the CAMB code (MGCAMB, see \cite{alessandra})
we have therefore computed the angular spectra for different values of $\lambda_1^2$.

\begin{figure}[h!]
\centerline{\includegraphics[width=9cm]{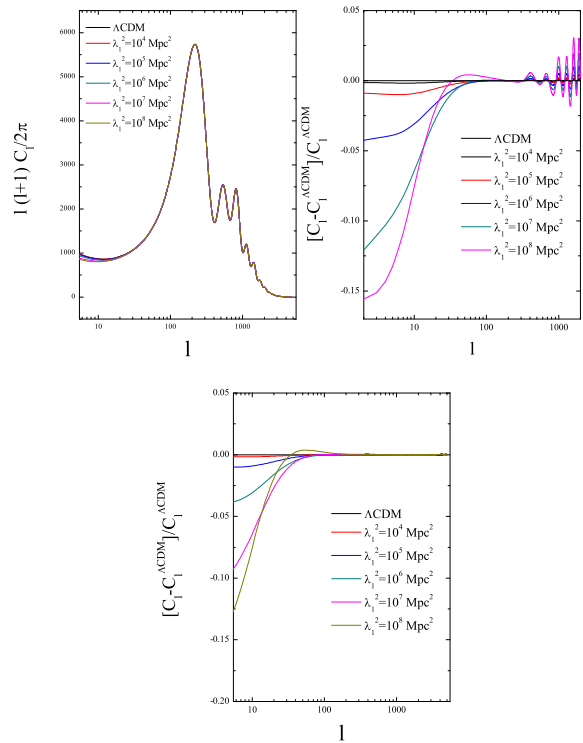}}
\caption{\label{fig:merge}CMB anisotropy power spectra (Top Left Panel) in a modified gravity model that mimics
the background expansion of $\Lambda$-CDM. As we can see (Top Right Panel) the main differences arise on large angular
scales due to the Integrated Sachs Wolfe effect and on small angular scales due to lensing. If lensing is not considered
the spectra are identical on small scales (Bottom Panel).}
\end{figure}

In Figure \ref{fig:merge} we plot the $C_l$ temperature anisotropy spectrum
for different values of $\lambda_1$, including $\lambda_1=0$ reproducing
the $\Lambda$CDM model; we also plot the relative difference between modified $C_l$ and the $\Lambda$CDM model. It is
possible to notice that while differences arise for large scale due to the ISW effect
also differences are well present at small angular scales due to the modification in the lensing potential.

\begin{figure}[h!]
\centerline{\includegraphics[width=8cm]{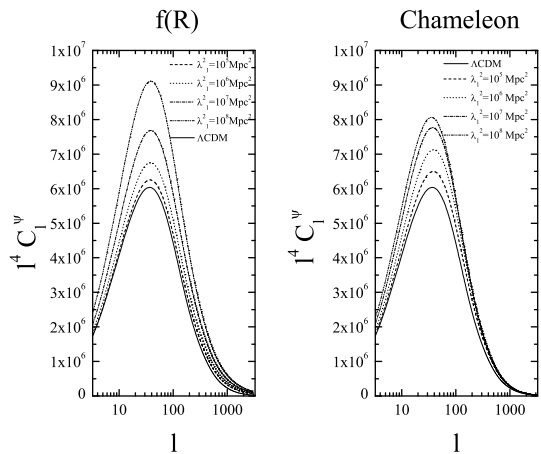}}
\caption{\label{fig:potgm}Lensing potential for different $f(R)$ models (left panel) and
Chameleon models (right panel) in function of different choices of the compton length $\lambda_1$.}
\end{figure}

In Figure \ref{fig:potgm}, left panel, we plot the corresponding lensing potential spectra.
As we can see, modified gravity could
enhance the lensing signal in a substantial way, and with greater effects respect to $c_s^2$.
One should also bear in mind that the models considered mimic a cosmological constant for the
background evolution. CMB lensing could therefore be a powerful method to disentangle modified gravity
from a cosmological constant. Values of  $\lambda_1^2 >10^6\ {\rm Mpc}^2$ could be ruled out by solar system tests even for
$f(R)$ theories that shows a chameleon mechanism; for example we can relate $\lambda_1$ to the parameter
$f_{R0}$ of \cite{HS} obtaining
$$f_{R0}=\frac{\lambda_1^2H_0^2(4-3\Omega_m)}{-1-n+\lambda_1^2H_0^2(-4+3\Omega_m)}$$
which leads, for the chosen value of $\lambda_1^2$, to a $|f_{R0}|\sim0.1$ which is barely compatible with
solar system bounds.

Also in Figure \ref{fig:potgm}, right panel, we plot the lensing potentials for a Chamelon
scalar-tensor theory (see \cite{alessandra}). This model can be obtained by choosing $\beta_1=9/7$, $\beta_2=7/9$ and $s=2$.
As we can see, again, modified gravity enhances the lensing potential even if the dependence
from $\lambda_1$ is different.

\section{Analysis Method}

In order to analyze the impact of a non-standard weak lensing component on
parameters inference we simply parameterize the weak lensing signal by defining a fudge
scaling parameter affecting the lensing potential power spectrum
(see \cite{ermi08}):
\begin{equation}
  \label{eq:wel}
  C_\ell^{\psi} \rightarrow A_L C_\ell^{\psi}.
\end{equation}

In other words, parameter $A_L$ effectively multiplies the matter
power lensing the CMB by a known factor. $A_L=0$ is therefore
equivalent to a theory that ignores lensing of the CMB, while $A_L=1$
gives the standard lensed theory and $A_L>1$ may suggest modified gravity or $c_s^2<1$.
 Since at the scales of interest the main effect of lensing is purely to smooth
the peaks in the data, $A_L$ can also be seen as a fudge parameter controlling the amount of
smoothing of the peaks (see \cite{ermi08}).

\begin{figure}[h!]
\centerline{\includegraphics[width=8cm]{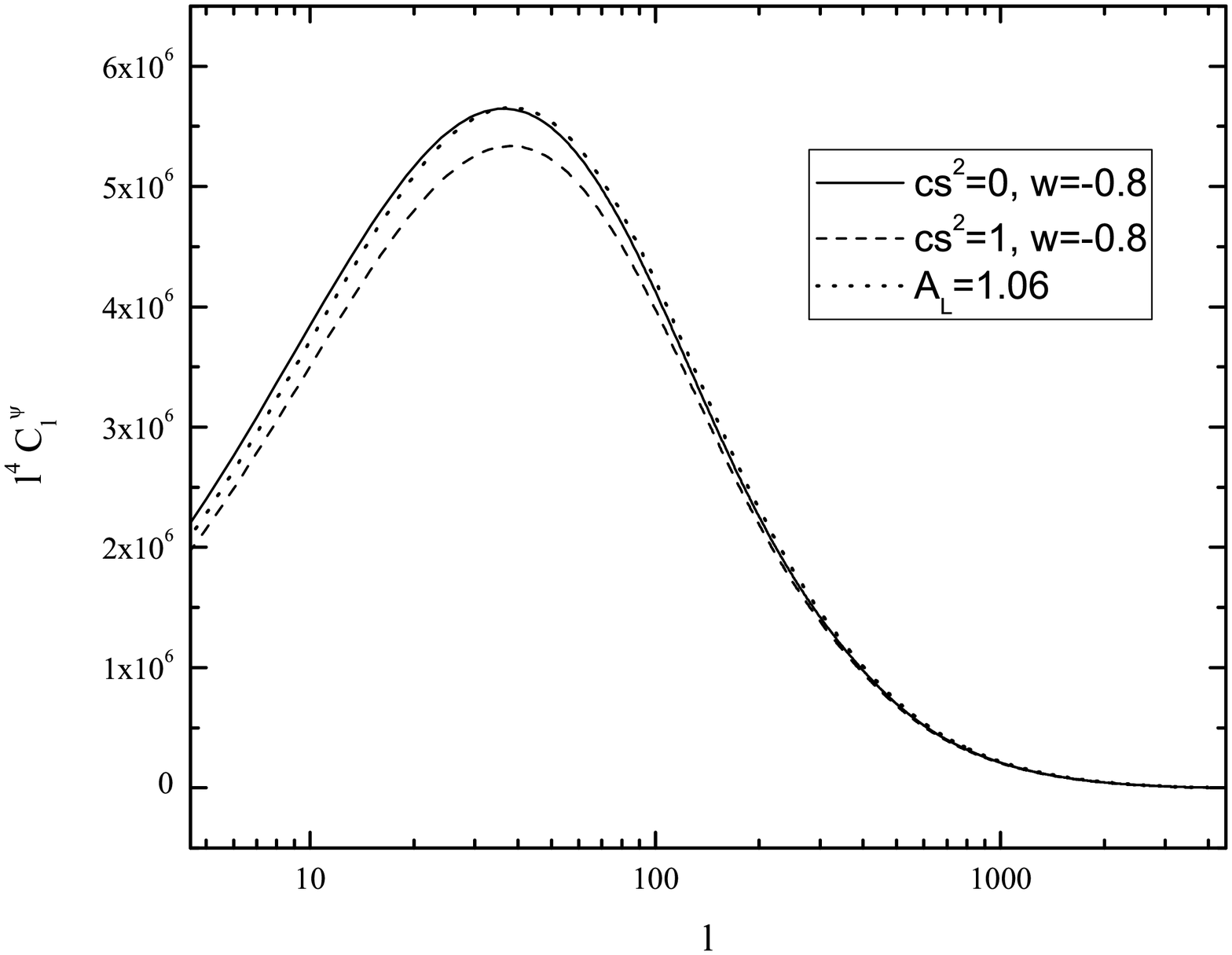}}
\centerline{\includegraphics[width=8cm]{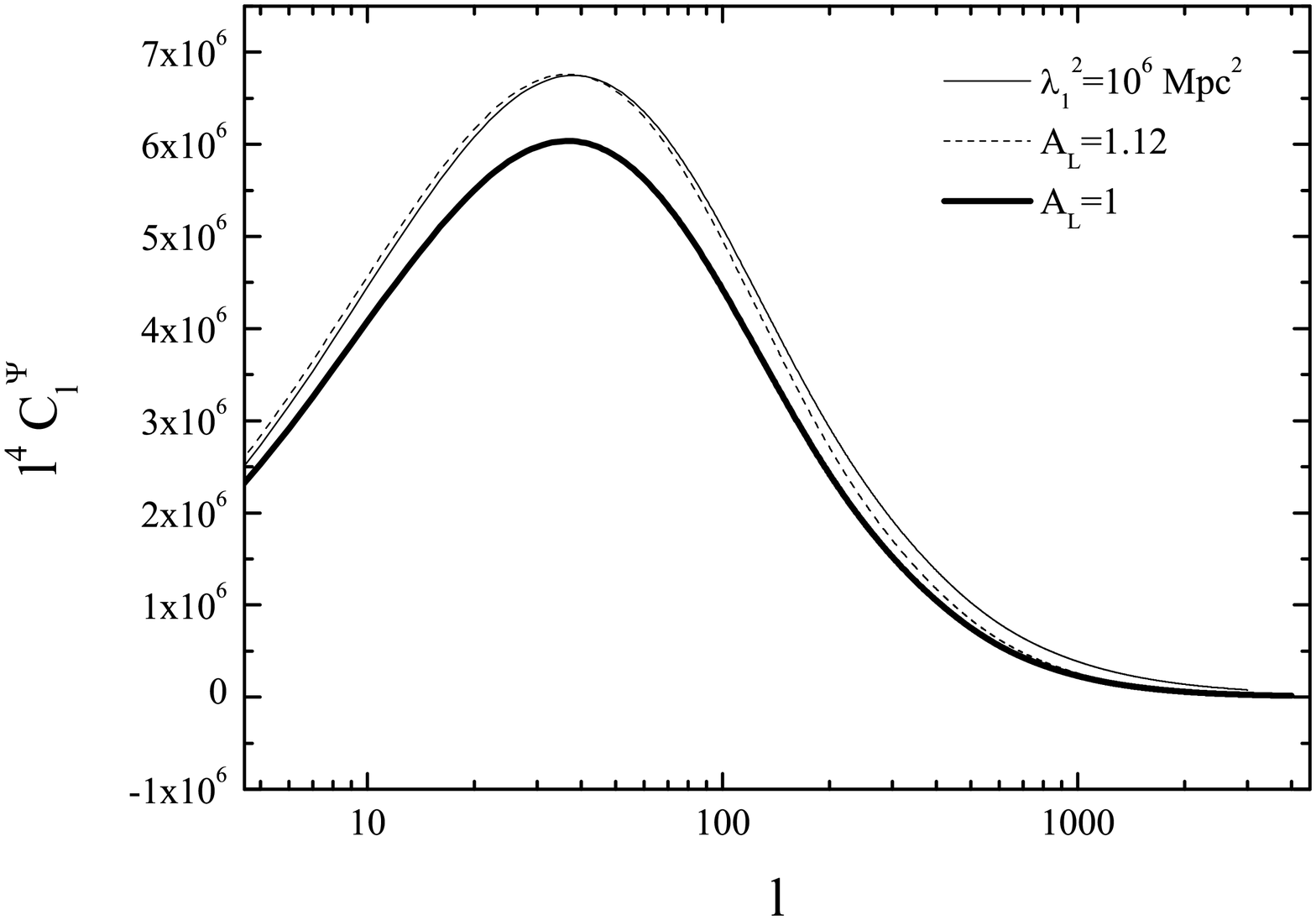}}
\caption{\label{fig:altut}Lensing potential for different dark energy (Top Panel) and modified gravity (Bottom Panel)
models. The $A_L$ parameter can well approximate the higher amplitude of the potential and the
deviations from the standard case.}
\end{figure}

As we can see in Figure \ref{fig:altut} the inclusion of this parameter can well recover
a non standard weak lensing signal. This is in agreement with the results
presented in \cite{shk06} where a one-parameter description of lensing could reasonably
recover all the main information achievable from CMB lensed spectra.

In particular we found that for $w=-0.8$
a model with $c_s=1$ and $A_L=1.06$ can perfectly reproduce the same
scenario with $c_s^2=0$, while a modified gravity model, compatible with
solar system bounds, can be approximated by $A_L=1.12$.

In what follows we provide constraints on $A_L$ by analyzing simulated datasets for future experiments
considering also B polarization modes\footnote{We ignored primordial tensor modes}.
Moreover, we investigate the impact of a wrong assumption related to  the
weak lensing projected potential amplitude on the determination of the cosmological
parameters.

We constrain the $A_L$ by a COSMOMC analysis of future mock datasets.
The analysis method we adopt is based on the
publicly available Markov Chain Monte Carlo package \texttt{cosmomc}
\cite{Lewis:2002ah} with a convergence diagnostic  done through the
Gelman and Rubin statistic.  We sample the following
eight-dimensional set of cosmological parameters, adopting flat priors
on them: the baryon and cold dark matter densities $\omega_{\rm b}$ and
$\omega_{\rm c}$, the ratio of the sound horizon to the angular diameter
distance at decoupling, $\theta_s$, the scalar spectral index $n_S$,
the overall normalization of the spectrum $A$ at $k=0.05$ {\rm Mpc}$^{-1}$,
the optical depth to reionization, $\tau$. Furthermore, we consider
purely adiabatic initial conditions and we impose spatial flatness.

We created several full mock datasets (temperature, E and B polarization modes)
 with noise properties consistent
with Planck \cite{:2006uk}, EPIC (\cite{Baumann:2008aq} and \cite{epic}),
and for SPT \cite{spt} (see Table \ref{tab:exp}),
assuming, as the fiducial model, the best-fit from the WMAP plus ACBAR
analysis of Ref.~\cite{wmap5} with
$\omega_{\rm b}=0.0227$, $\omega_{\rm c}= 0.113$, $n_S=0.0973$, $\tau=0.0908$
and $A_L=1$.

\begin{table}[!htb]
\begin{center}
\begin{tabular}{rccc}
Experiment & Channel & FWHM & $\Delta T/T$ \\
\hline
Planck & 143 & 7.1'& 2.2\\
$f_{sky}=1.0$\\
\hline
Epic 2m& 150 & 5' & 0.44 \\
$f_{sky}=0.85$ & & &\\
\hline
SPT & 150 & 1' & 10 \\
$f_{sky}=0.06$ & & &\\
\end{tabular}
\caption{Planck, EPIC, and SPT experimental specifications.  Channel frequency is given
in GHz, FWHM in arcminutes and noise per pixel in $10^{-6}$ for the Stokes-I parameter;
the corresponding sensitivities for the Stokes Q and U parameters are related to this by a factor of $\sqrt{2}$.}
\label{tab:exp}
\end{center}
\end{table}

We consider for each channel a detector noise of $w^{-1} = (\theta\sigma)^2$ where $\theta$ is the FWHM of
 the beam assuming Gaussian profile and $\sigma$ is the sensitivity $\Delta T/T$ both
from Table~\ref{tab:exp}. We therefore add to each $C_\ell$ fiducial spectra a noise
spectrum given by:
\begin{equation}
N_\ell = w^{-1}\exp(l(l+1)l_b^2) \, ,
\end{equation}
where $l_b$ is given by $l_b \equiv \sqrt{8\ln2}/\theta$.
We analyze these datasets with a full-sky exact likelihood routine as in Ref.~\cite{Lewis:2005tp}. In the case of
SPT we also include a $f_{sky}^2\sim0.06$ pre-factor ($f_{sky}=1$ for Planck and $f_{sky}=0.85$ for EPIC).

\section{Results}

\begin{table}[h!]
\begin{center}
\begin{tabular}{lcc}
\hline
Experiment & Dataset & Limits on $A_L$\\
\hline
\vspace{0.2cm}
Planck & w/o Bmodes & $1.00^{+0.06+0.12}_{-0.06-0.11}$\\
\vspace{0.2cm}
Planck & with Bmodes & $1.00^{+0.06+0.12}_{-0.06-0.12}$\\
\vspace{0.2cm}
EPIC & w/o Bmodes & $1.000^{+0.013+0.027}_{-0.013-0.026}$\\
\vspace{0.2cm}
EPIC & with Bmodes & $1.000^{+0.011+0.024}_{-0.011-0.023}$\\
\vspace{0.2cm}
SPT & w/o Bmodes & $1.00^{+0.26+0.60}_{-0.26-0.44}$\\
\vspace{0.2cm}
SPT & with Bmodes & $1.00^{+0.09+0.42}_{-0.13-0.34}$\\
\end{tabular}
\caption{Fiducial Model parameters and limits on $A_L$ for Planck and EPIC satellites and for the
SPT experiment, We report errors at $68\%$ and $95\%$ confidence level.}
\label{tab:al}
\end{center}
\end{table}

In Table~\ref{tab:al} we report the constraints on the $A_L$ parameter achievable by future
experiments. Planck will reach an accuracy of about $\sim 6 \%$ at
$1 \sigma$ level while EPIC will constrain $A_L$ at $\sim 1 \%$ level (always $1 \sigma$).
It is interesting to note that the exclusion of the B-mode polarization channel from
these experiments has small effects on the determination of $A_L$. In a few words, the
smearing of temperature and polarization anisotropies induced by lensing
provides a larger statistical evidence for $A_L$ than a detection of lensed B-mode polarization.

This will not be the case for the SPT experiment, where the
higher angular resolution will permit better measurement of the B-mode polarizations signal.
In the SPT case, the inclusion of $B$ modes practically halves the error bars on $A_L$.
However, as already discussed in \cite{shk04} the sample variance on the small scale B modes
could be larger by a factor of $\sim 10$, reflecting the variance of the larger scale lenses that generate them.
This degradation effect should be considered for SPT and could strongly affect the reported results
on $A_L$ when B-modes are considered. Being the degradation effect much smaller for the
$TT$, $EE$ and $TE$ modes, in what follows we only consider forecasts for the Planck and
EPIC experiments.

Comparing the future constraints on $A_L$ with the expected signal from dark energy perturbations,
it is clear that the Epic experiment may have enough sensitivity
to constraint the sound speed parameter $c_s^2$ if $w \sim -0.8$.
In order to test this, we have investigated the constraints
achievable by Planck and EPIC on the $c_s^2-w$ plane under the assumption of a fiducial model
with $w=-0.8$, $c_s^2=0$ and the remaining parameters fixed as above.

\begin{figure}[h!]
 \centerline{\includegraphics[width=7cm]{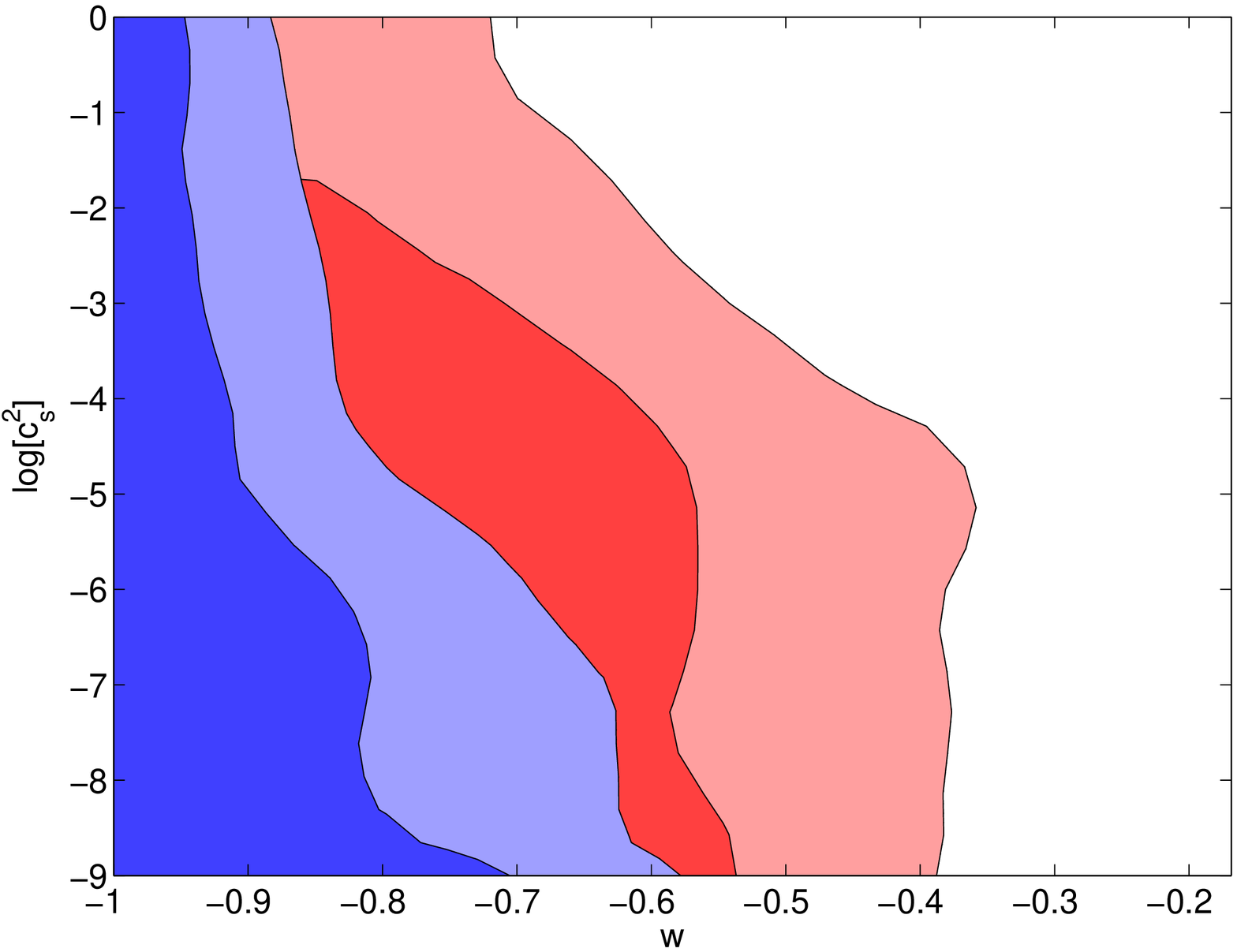}}
 \centerline{\includegraphics[width=7cm]{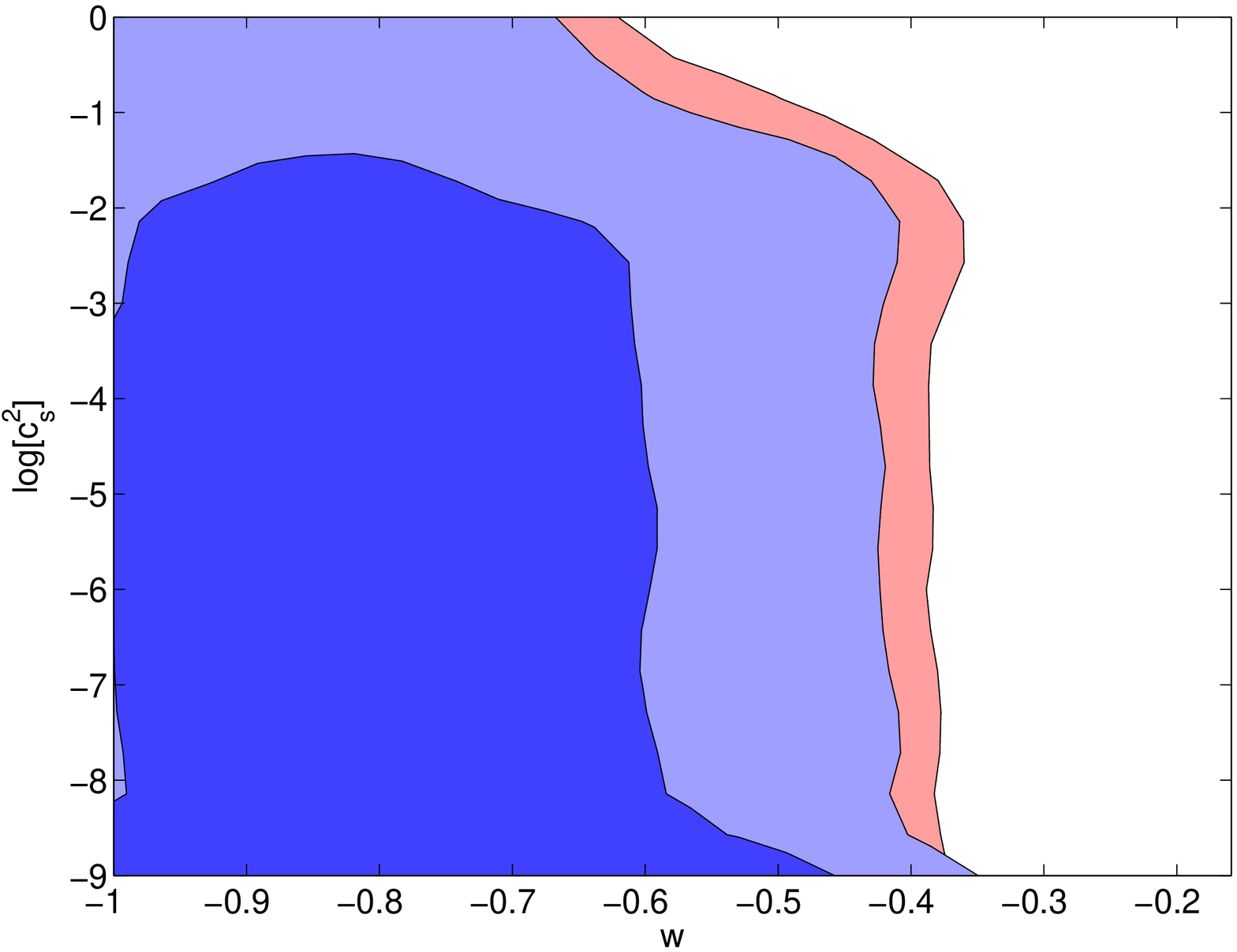}}
 \centerline{\includegraphics[width=7cm]{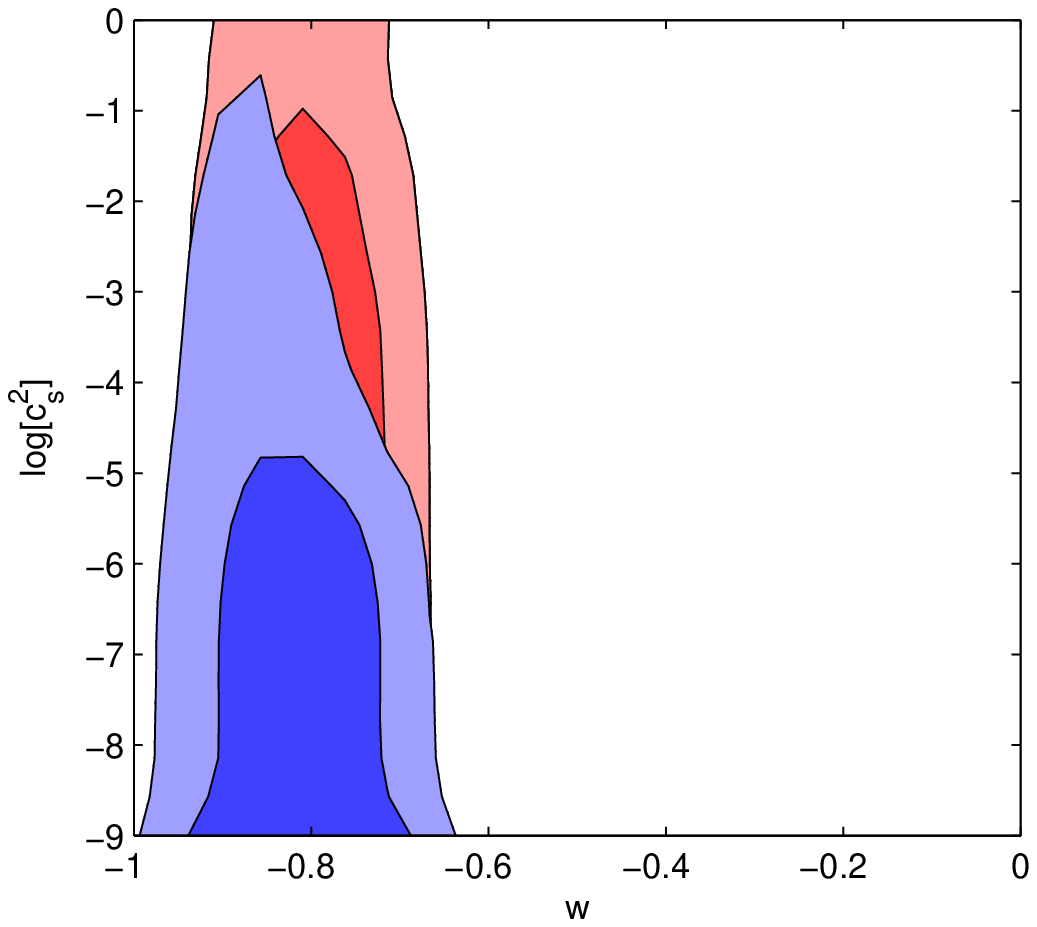}}
\caption{\label{cs2-w} $68 \%$ and $95 \%$ c.l. constraints on the $c_s^2-w$ plane
from Planck (Red) and EPIC (Blue) when lensing is considered (Top Panel) and without considering
lensing (Middle Panel). The fiducial model assumes $w=-0.8$ and $c_s^2=0$.
In the Bottom Panel we can see the effect of adding an external prior on
$w=-0.8\pm0.05$ at $68 $ c.l.. A combined analysis with CMB data could
discriminate values of $c_s^2$ lower than one.}
\end{figure}

In Figure \ref{cs2-w} we plot the $68 \%$ and $95 \%$ confidence levels expected from
Planck and EPIC with and without considering the lensing signal. As we can see,
when lensing is considered, $c_s^2$ is strongly constrained for larger values
of $w$. Adding complementary information on $w$ such that $w=-0.8\pm0.1$ yields
$c_s^2<0.1$ from EPIC at $95\%$ c.l.

On the other hand, the constraints obtained on $A_L$ will probe modified gravity
scenarios. In particular, in the case of $f(R)$ models, an upper limit of $A_L<1.06$
from Planck ($A_L<1.01$ from EPIC) will bound the Compton wavelength to $\lambda_1^2 <2 \cdot 10^5 {\rm Mpc}^2$
(Planck) and $\lambda_1^2< 2 \cdot 10^4 {\rm Mpc}^2$ (EPIC). Models that are consistent with solar system
test can be significantly ruled out by EPIC. For Chameleons models the bound will be
similar with $\lambda_1^2 <8 \cdot 10^4 {\rm Mpc}^2$ (Planck) and $\lambda_1^2< 9 \cdot 10^3 {\rm Mpc}^2$ (EPIC).

The accuracy reachable by Planck and EPIC will therefore open the possibility of testing
non standard modification to gravity at cosmological length scales (\cite{serra09}).
It is interesting to quantify the impact of a wrong assumption
in the lensing potential on the constraints derivable from those experiments
and the corresponding ability of recovering the correct solution.

\begin{figure}[h!]
\includegraphics[width=4.2cm]{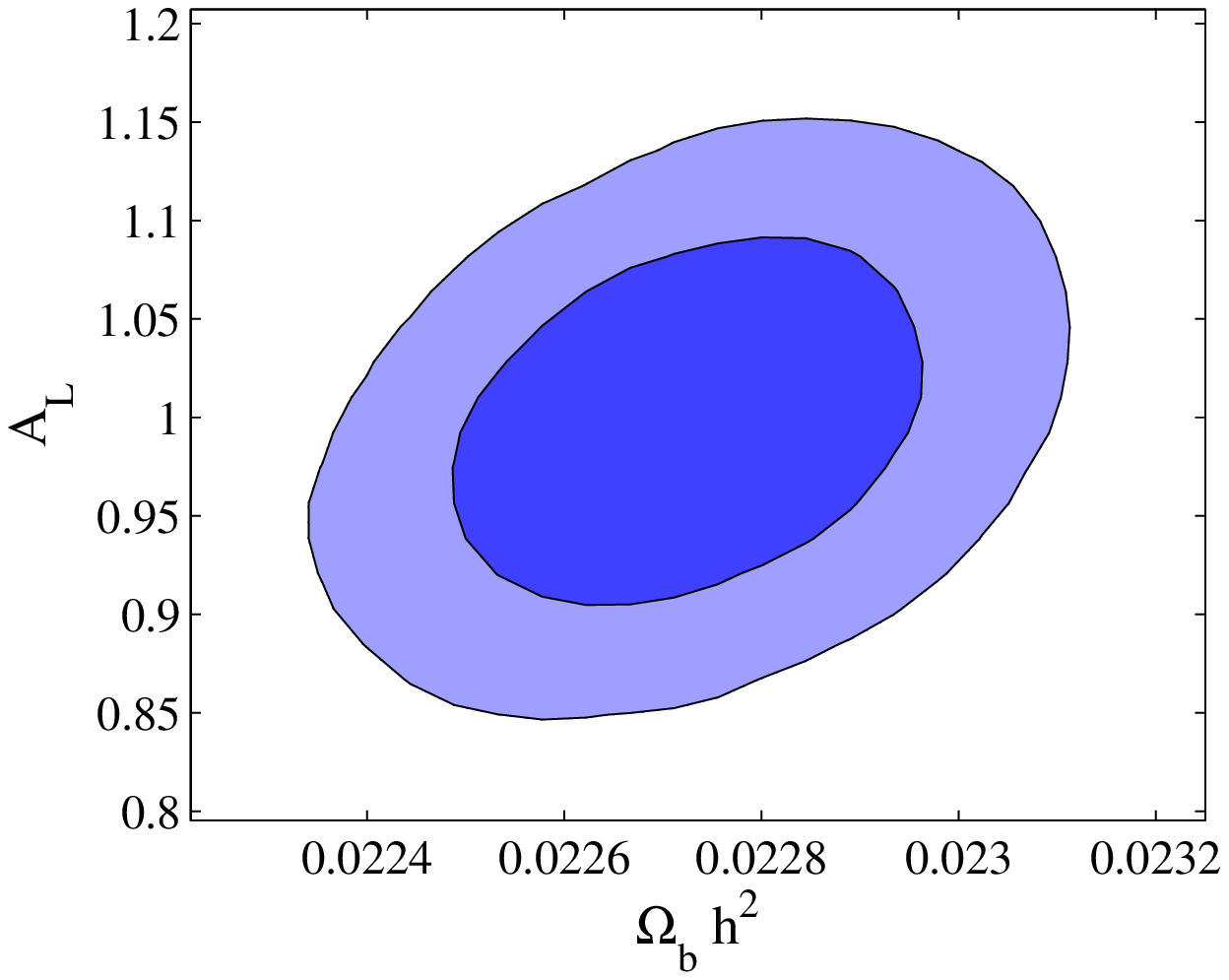}
\includegraphics[width=4.2cm]{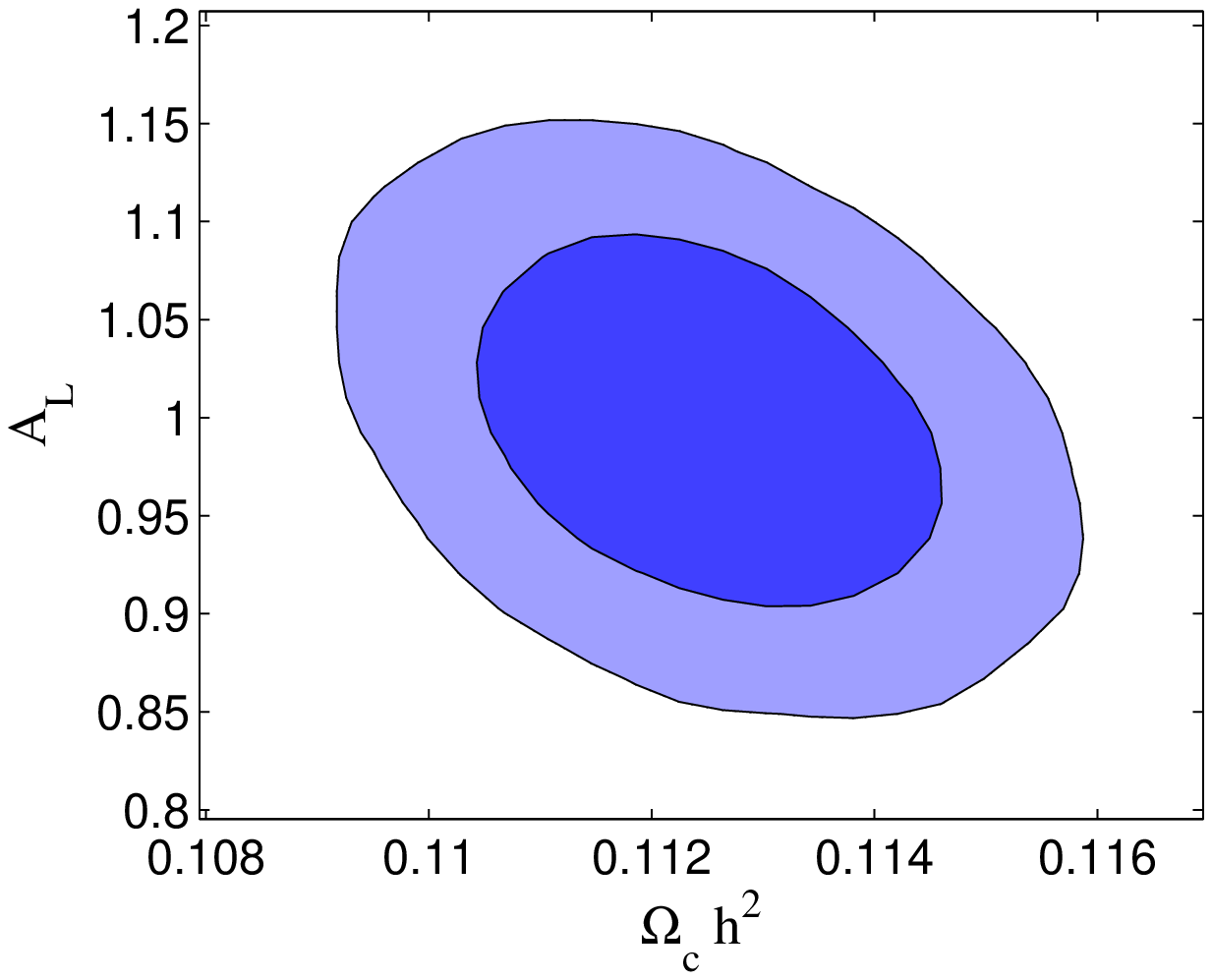}
\includegraphics[width=4.2cm]{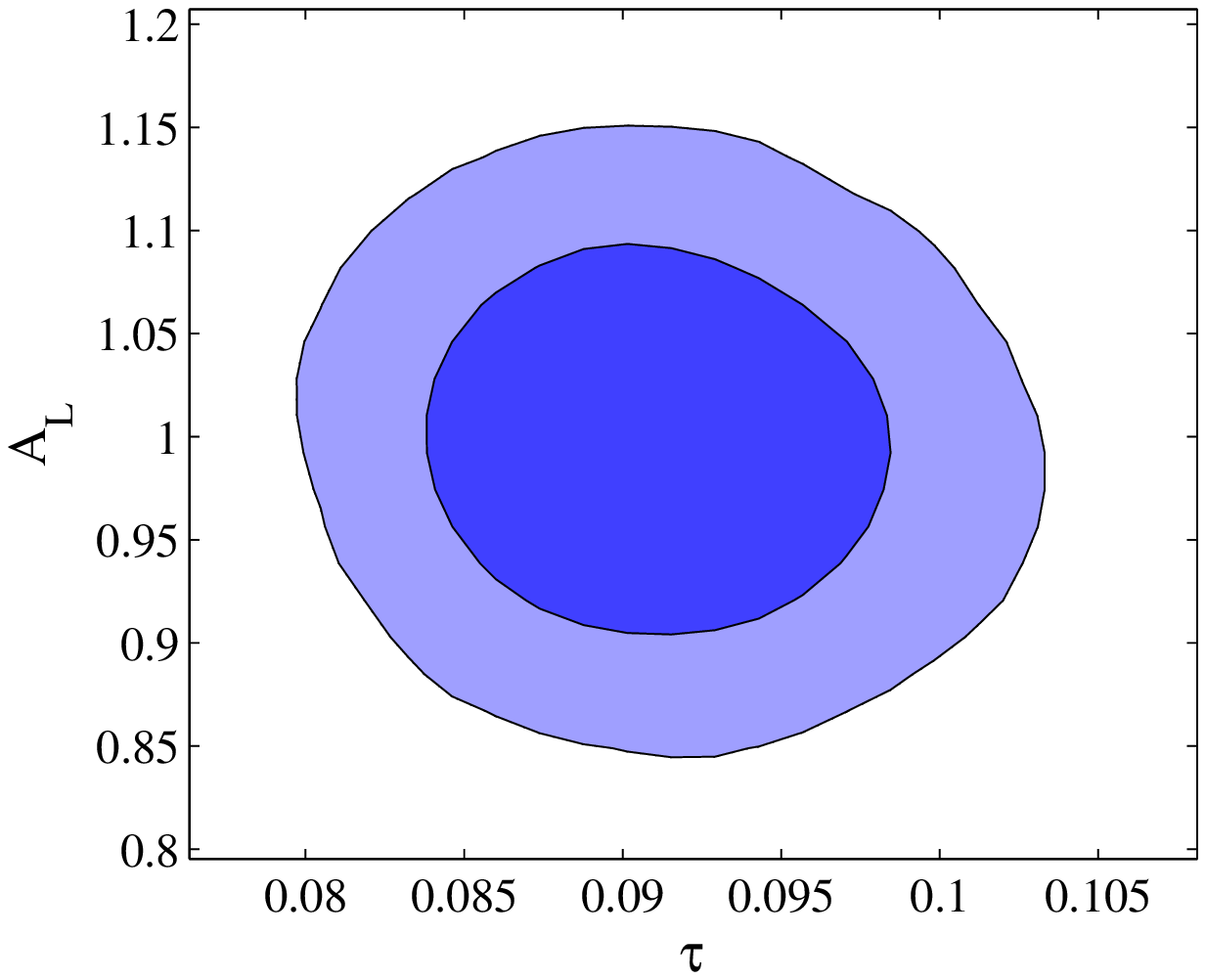}
\includegraphics[width=4.2cm]{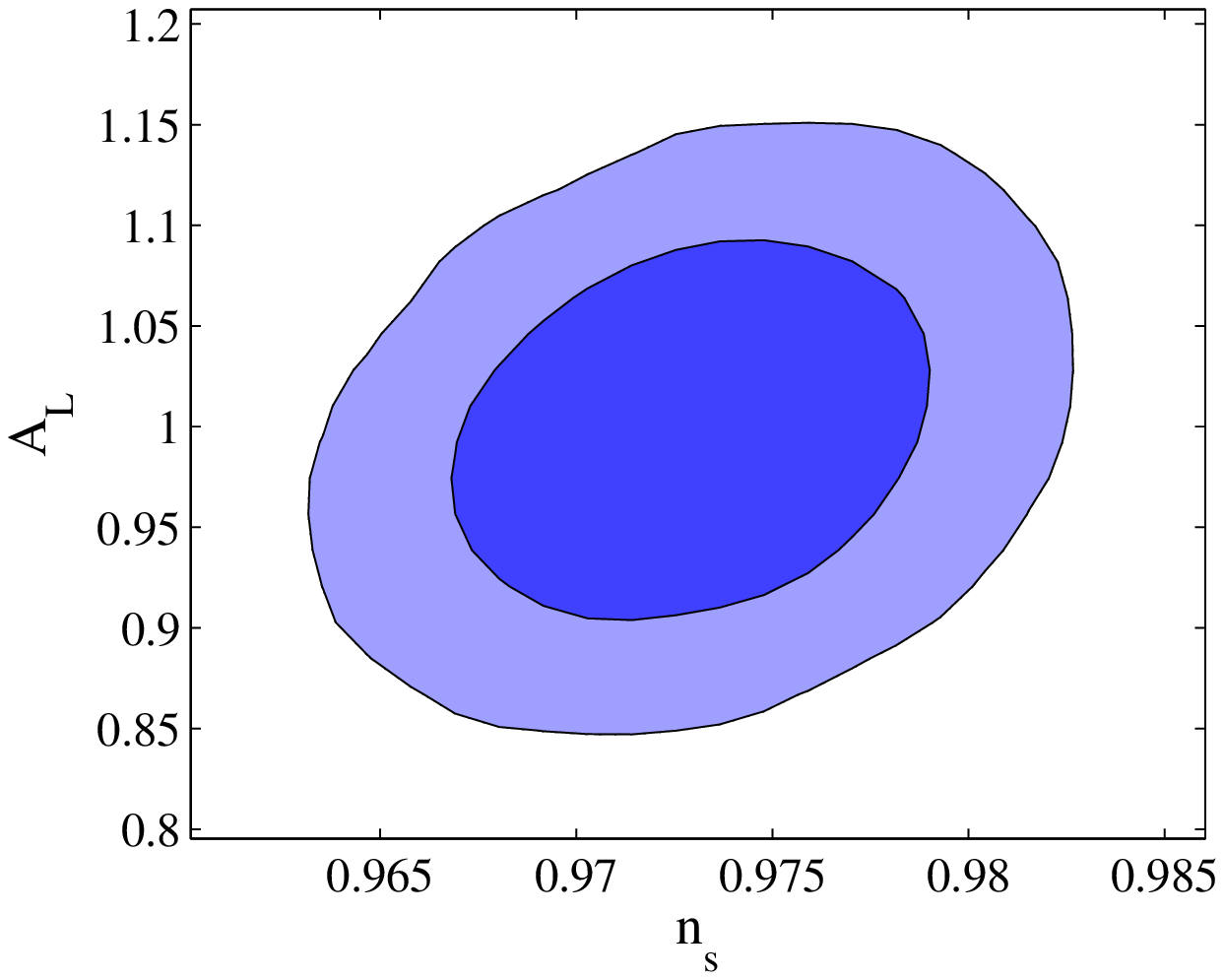}
\caption{\label{fig:deg_planck}$68 \%$ and $95 \%$ c.l. $2$-D Likelihood contour plots for $A_L$ versus different
cosmological parameters from the Planck experiment.}
\end{figure}

\begin{figure}[h!]
\includegraphics[width=4.2cm]{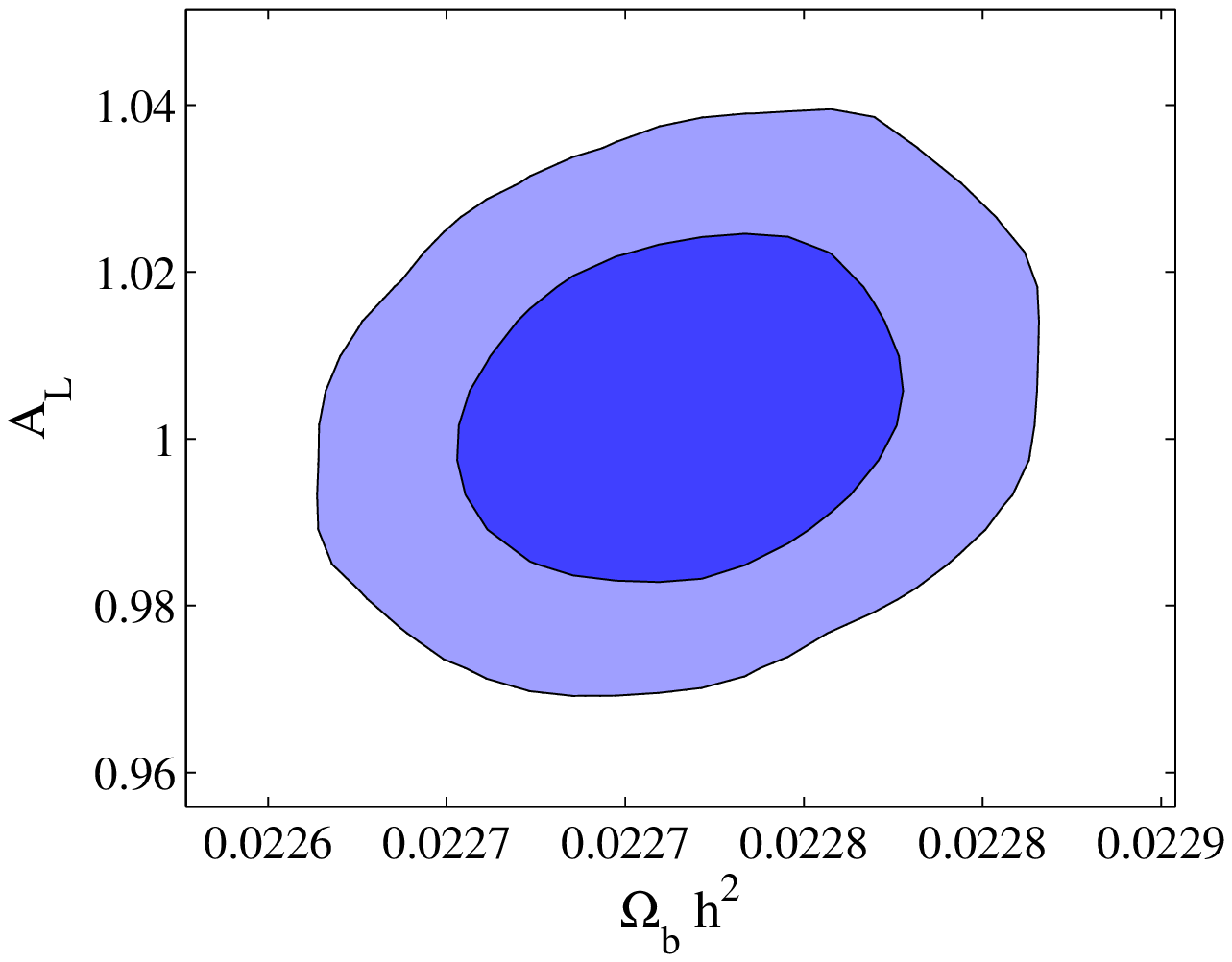}
\includegraphics[width=4.2cm]{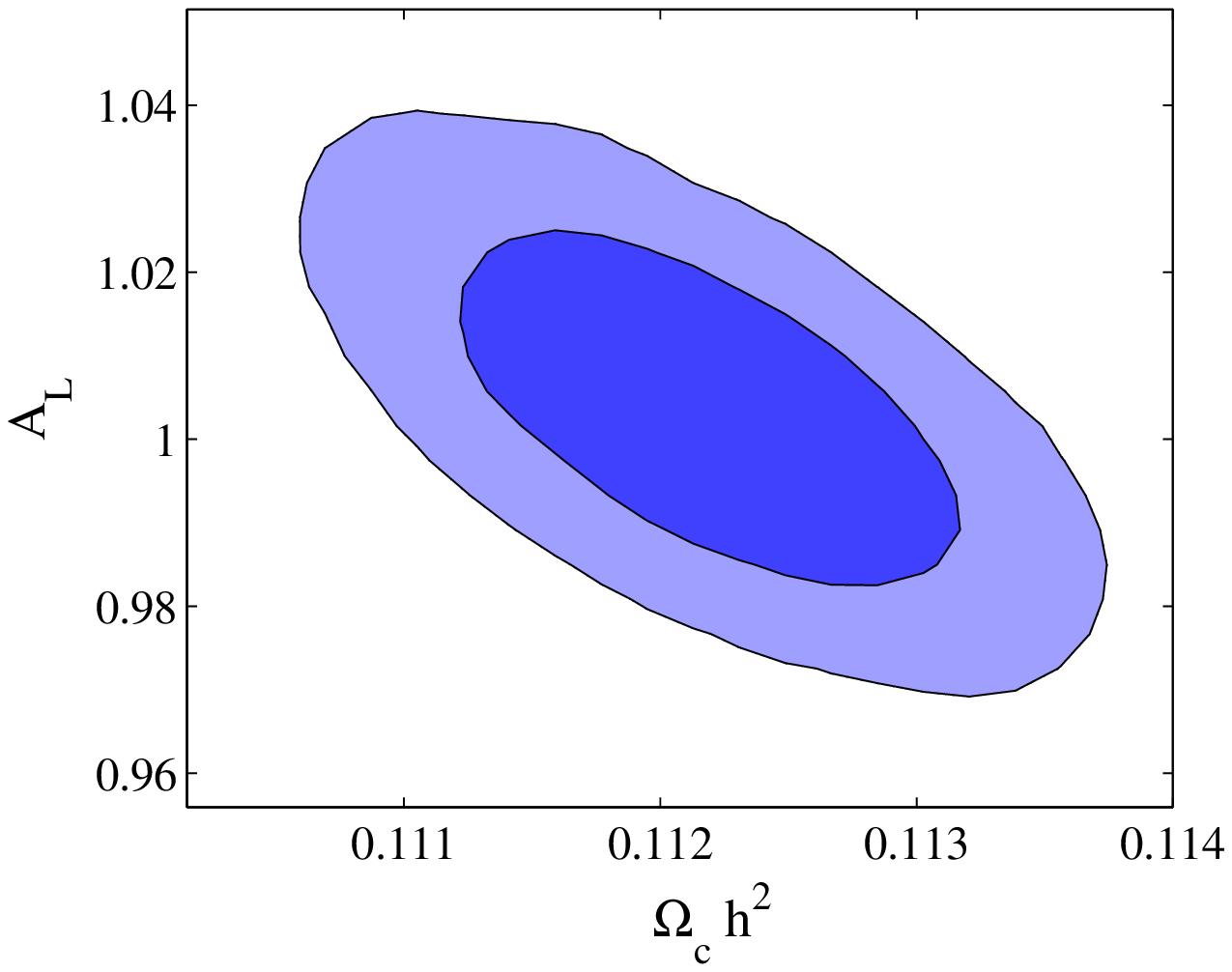}
\includegraphics[width=4.2cm]{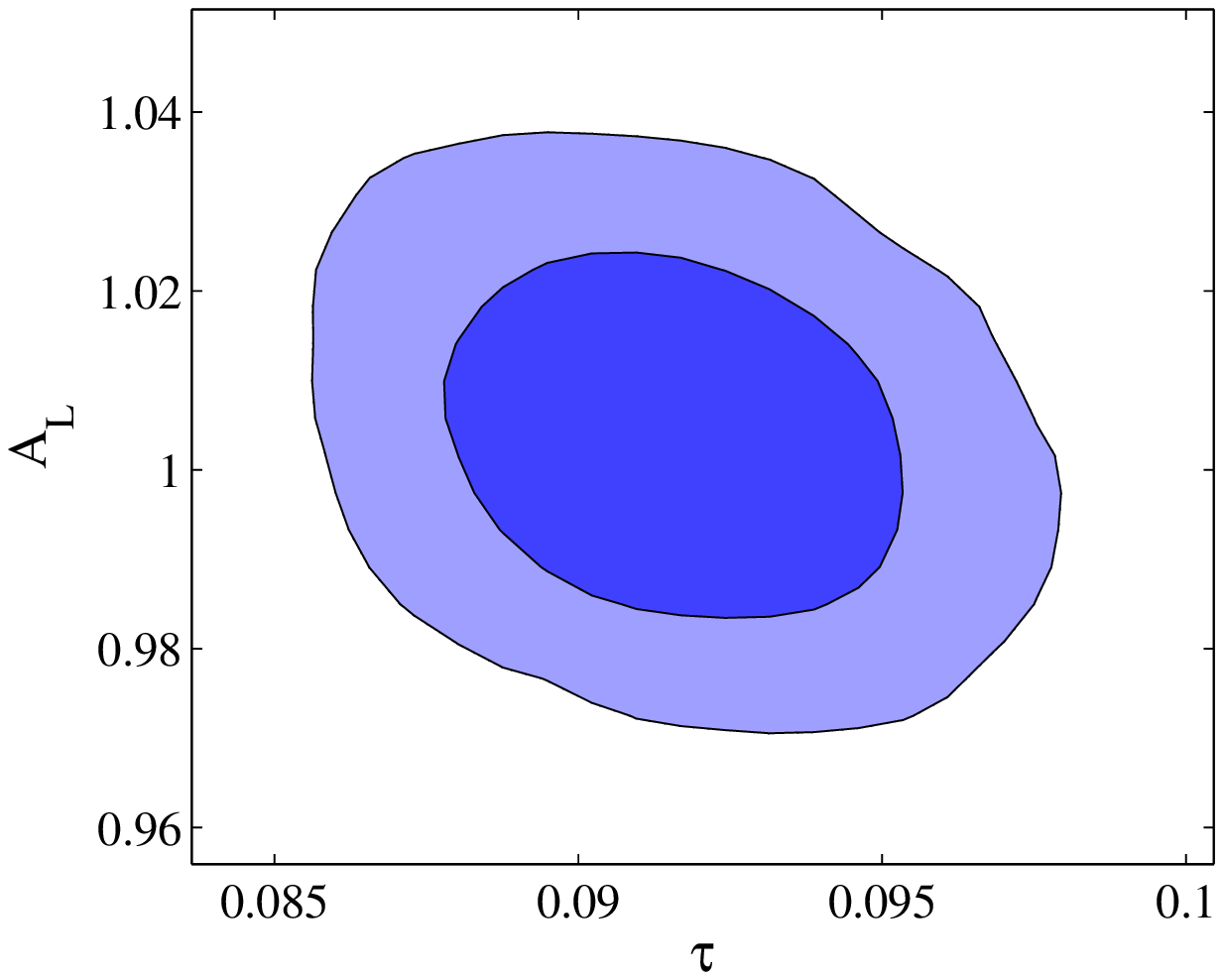}
\includegraphics[width=4.2cm]{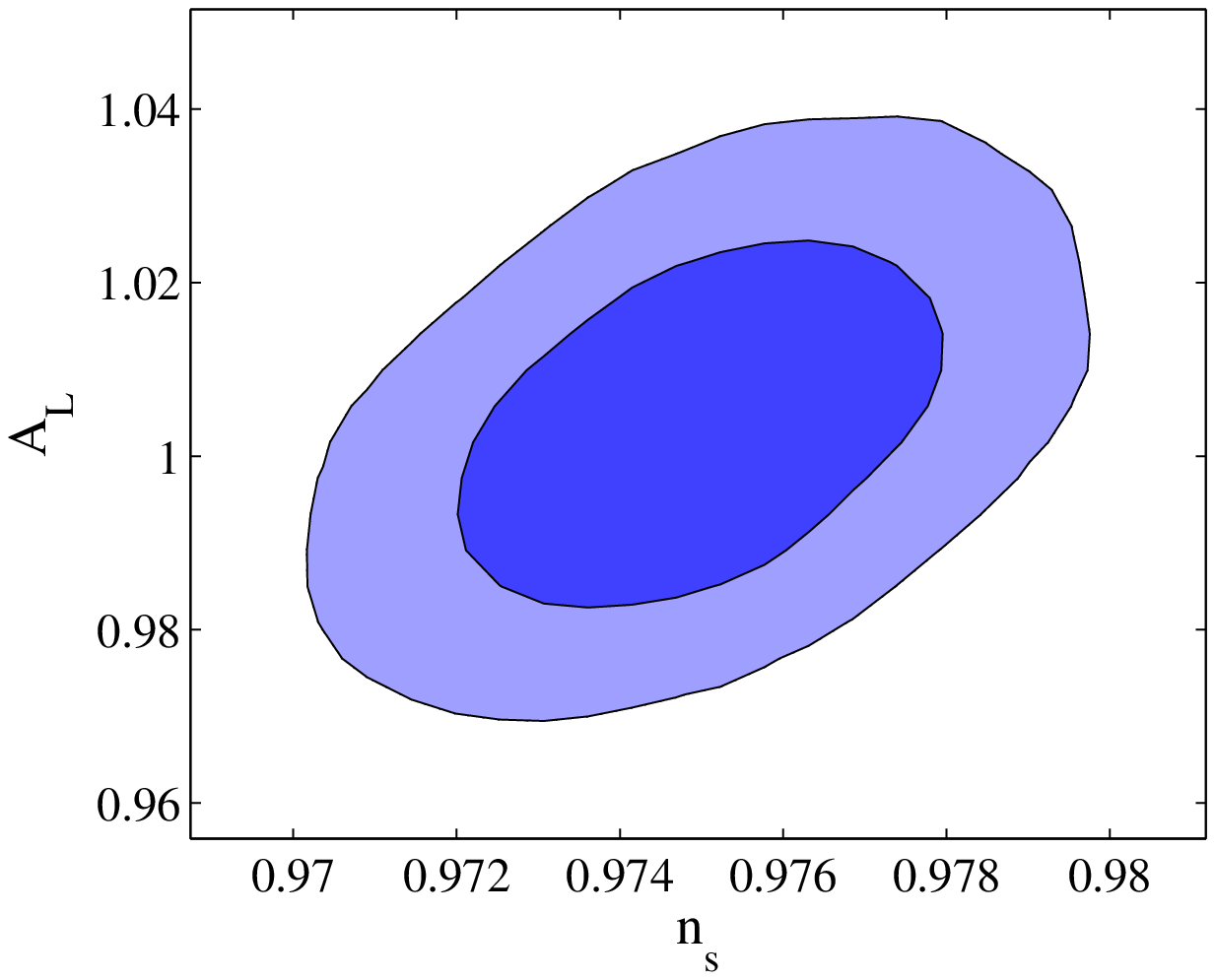}
\caption{\label{fig:deg_epic}$68 \%$ and $95 \%$ c.l. $2$-D Likelihood contour plots for $A_L$ versus different
cosmological parameters from the EPIC experiment.}
\end{figure}

In Figures \ref{fig:deg_planck} and \ref{fig:deg_epic} we can see the $2$-D likelihood contour plots
which show the degeneracies between the lensing parameter $A_L$ and the remaining cosmological parameters.
As we can see, a degeneracy appears between $A_L$ and the matter density $\omega_c$ and the
spectral index $n_S$ while a milder degeneracy is also present with the baryon density and the optical
depth $\tau$.  A larger value of $A_L$ will make larger values of $\omega_b$ and $n_S$
and smaller values of $\omega_c$ and $\tau$ compatible with the CMB data respect to the case with
$A_L=1$.

Including an uncertainty on $A_L$ will therefore relax the bound on these parameters,
especially on $n_S$ and $\omega_b$. It is however interesting to quantify the
error one could make if not including a marginalization over $A_L$ when analyzing future datasets.

We have therefore changed the fiducial model to $A_L=1.3$ for Planck ($A_L=1.05$ for EPIC) and re-analyzed
the mock datasets either with the wrong assumption of $A_L=1$, either letting $A_L$ to vary,

\begin{figure}[h!]
 \centerline{\includegraphics[width=9cm]{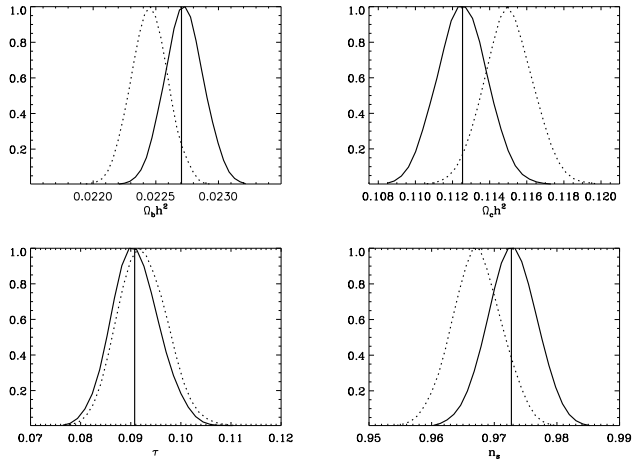}}
\caption{\label{fig:pk_bias}Bias introduced in cosmic parameters inference by the assumption of
a wrong weak lensing amplitude in the case of the Planck experiment. The fiducial model
has $A_L=1.3$. The solid line show the constraints when the $A_l$ parameter is let to vary,
while the dotted constraints are assuming $A_L=1$. The straight black line identifies the
fiducial model.}
\end{figure}

\begin{figure}[h!]
 \centerline{\includegraphics[width=9cm]{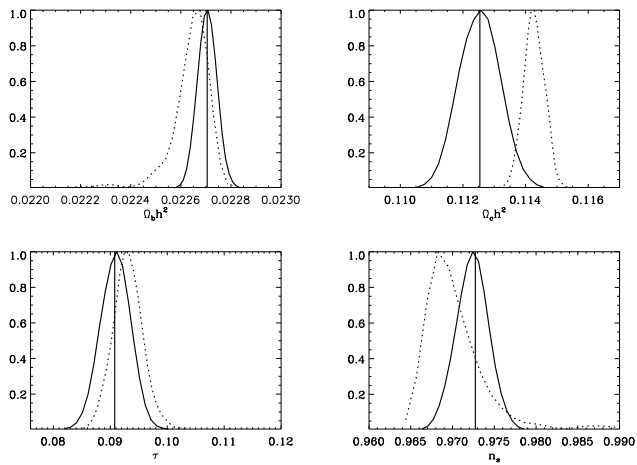}}
\caption{\label{fig:epic_bias}Bias introduced in cosmic parameters inference by the assumption of
a wrong weak lensing amplitude in the case of the EPIC experiment. The fiducial model
has $A_L=1.05$. The solid line show the constraints when the $A_l$ parameter is let to vary,
while the dotted constraints are assuming $A_L=1$. The straight black line identifies the
fiducial model.}
\end{figure}

As we can see from Figures \ref{fig:pk_bias} and \ref{fig:epic_bias}, not accounting for an higher
amplitude of the lensing potential could lead to misleading constraints on the baryon density,
the cold dark matter density the spectral index and the optical depth.
In particular, if we don't marginalize over $A_L$ we obtain the constraints $\omega_b=0.02249\pm0.00014$,
 $\omega_c=0.1150\pm0.0013$, $n_s=0.9672\pm0.0038$, $\tau =0.0924\pm0.0047$ for Planck and $\omega_b=0.02264\pm0.00006$,
 $\omega_c=0.11426\pm0.00034$, $\tau =0.0931 \pm 0.0024$ and $n_s=0.9704\pm0.0027$ for EPIC to be compared with the
 fiducial values of $\omega_b=0.0227$, $\omega_c=0.113$, $n_s=0.973$, $\tau =0.0908$. When a marginalization
 over $A_L$ is included we recover the correct fiducial input parameters.

\section{Conclusions}

In this paper we have investigated the ability of future measurements of CMB temperature
and polarization angular spectra to identify a non-standard weak lensing signal.
We have found that the amplitude of the weak lensing potential will be measured with
a $\sim 6\%$ accuracy from Planck ($1 \%$ from EPIC) at $68 \%$ confidence level
This will let to test the dark energy sound speed parameter $c_S^2$  for
values of the equation of state different from $-1$. If, for example, the
dark energy of state will be constrained at $w=-0.80\pm0.05$ at $68 \%$ c.l.
then a value of $c_s^2=0$ could be discriminated at more than $3$ standard deviations from
$c_s^2=1$ from an EPIC-like experiment.

Moreover, we have found that both Planck and EPIC could place new constraints on modified
gravity models. Translating the bounds on $A_L$ in the case of $f(R)$ models on
the Compton wavelength $\lambda_1$ we have
found that Planck will place a bound of $\lambda_1^2 <2 \cdot 10^5 {\rm Mpc}^2$
 while EPIC will reach $\lambda_1^2 < 2 \cdot 10^4 {\rm Mpc}^2$.
For Chameleons models the bound will be
similar with $\lambda_1^2 <8 \cdot 10^4 {\rm Mpc}^2$ (Planck) and $\lambda_1^2< 9 \cdot 10^3 {\rm Mpc}^2$ (EPIC).
This will present at least a two order of magnitude improvement on the constraint
$\lambda_1^2 < 4 \cdot 10^7 {\rm Mpc}^2$ from current CMB data \cite{song07}.
Finally one should remember that constraints on $A_L$ can be placed by considering
higher order correlations and extracting the lensing deflection field from the CMB maps.
Several methods have been proposed as the quadratic estimator method (see \cite{huokamoto})
and the iterative method by Hirata \& Seljak (\cite{hirataseljak}).
Estimating the noise in the case of a quadratic estimator, we have found that a $\sim 1 \%$ error
on $A_L$ could be reached by EPIC of about the same order of the constraint presented here
obtained from measurements of the angular power spectra.
Since experimental systematics and foregrounds may strongly affect future CMB measurements but 
affect the angular spectra and the quadratic estimator in a different way the two
methods are complementary and will both provide valuable information to the dark energy component.

\vspace{0.6cm}
\noindent {\bf Acknowledgment}\\
Research work of EC at UC Irvine was supported by NSF CAREER AST-0605427.
AM thanks the University of California Irvine for hospitality.
 AS acknowledges financial support from the Berkeley Center for Cosmological Physics.  This research
has been supported by ASI contract I/016/07/0 "COFIS".


\begin{thebibliography}{99}

\bibitem{wmap5}
  G.~Hinshaw {\it et al.}  [WMAP Collaboration],
  arXiv:0803.0732 [astro-ph];
  E.~Komatsu {\it et al.},
  arXiv:0803.0547 [astro-ph].

\bibitem{cbi}
  J.~L.~Sievers {\it et al.},
  Astrophys.\ J.\  {\bf 660} (2007) 976
  [arXiv:astro-ph/0509203].


\bibitem{acbar}
  C.~L.~Reichardt {\it et al.},
  arXiv:0801.1491 [astro-ph].

\bibitem{boom03}
  W.~C.~Jones {\it et al.},
  arXiv:astro-ph/0507494;  F.~Piacentini {\it et al.},
  arXiv:astro-ph/0507507;
  arXiv:astro-ph/0507514.

\bibitem{planck}
    [Planck Collaboration],
  arXiv:astro-ph/0604069.

\bibitem{spt}
  J.~E.~Ruhl {\it et al.}  [The SPT Collaboration],
  Proc.\ SPIE Int.\ Soc.\ Opt.\ Eng.\  {\bf 5498}, 11 (2004)
  [arXiv:astro-ph/0411122].



\bibitem{spider}
  B.~P.~Crill {\it et al.},
  arXiv:0807.1548 [astro-ph].

\bibitem{ebex}
  P.~Oxley {\it et al.},
  Proc.\ SPIE Int.\ Soc.\ Opt.\ Eng.\  {\bf 5543} (2004) 320
  [arXiv:astro-ph/0501111].

\bibitem{epic}
  J.~Bock {\it et al.},
  arXiv:0805.4207 [astro-ph];
  J.~Bock {\it et al.}  [EPIC Collaboration and EPIC Collaboration and EPIC
                  Collaboration and  ],
  arXiv:0906.1188 [astro-ph.CO].

\bibitem{cmbfast}
  U.~Seljak and M.~Zaldarriaga,
  Astrophys.\ J.\  {\bf 469} (1996) 437
  [arXiv:astro-ph/9603033].

\bibitem{camb}
  A.~Lewis, A.~Challinor and A.~Lasenby,
  Astrophys.\ J.\  {\bf 538} (2000) 473
  [arXiv:astro-ph/9911177].

\bibitem{accuracy}
  U.~Seljak, N.~Sugiyama, M.~J.~.~White and M.~Zaldarriaga,
  Phys.\ Rev.\  D {\bf 68} (2003) 083507
  [arXiv:astro-ph/0306052].


\bibitem{hirata}
  C.~M.~Hirata,
  Phys.\ Rev.\  D {\bf 78} (2008) 023001
  [arXiv:0803.0808 [astro-ph]];
  arXiv:0904.2220 [astro-ph.CO].

\bibitem{bean}
  G.~Efstathiou and J.~R.~Bond,
  Mon.\ Not.\ Roy.\ Astron.\ Soc.\  {\bf 304} (1999) 75
  [arXiv:astro-ph/9807103];
  R.~Bean and A.~Melchiorri,
  Phys.\ Rev.\  D {\bf 65} (2002) 041302
  [arXiv:astro-ph/0110472].

\bibitem{kowalski}
  M.~Kowalski {\it et al.}  [Supernova Cosmology Project Collaboration],
  Astrophys.\ J.\  {\bf 686} (2008) 749
  [arXiv:0804.4142 [astro-ph]].

\bibitem{silvestri}
  A.~Silvestri and M.~Trodden,
  arXiv:0904.0024 [astro-ph.CO].

\bibitem{marcald}
  R.~R.~Caldwell and M.~Kamionkowski,
  arXiv:0903.0866 [astro-ph.CO].

\bibitem{isw}
  R.~K.~Sachs and A.~M.~Wolfe,
  Astrophys.\ J.\  {\bf 147} (1967) 73
  [Gen.\ Rel.\ Grav.\  {\bf 39} (2007) 1929].

\bibitem{song07}
  Y.~S.~Song, H.~Peiris and W.~Hu,
  Phys.\ Rev.\  D {\bf 76} (2007) 063517
  [arXiv:0706.2399 [astro-ph]].

\bibitem{caldwell}
  R.~Caldwell, A.~Cooray and A.~Melchiorri,
  Phys.\ Rev.\  D {\bf 76}, 023507 (2007)
  [arXiv:astro-ph/0703375];
 S.~F.~Daniel, R.~R.~Caldwell, A.~Cooray and A.~Melchiorri,
  Phys.\ Rev.\  D {\bf 77} (2008) 103513
  [arXiv:0802.1068 [astro-ph]];
  S.~F.~Daniel, R.~R.~Caldwell, A.~Cooray, P.~Serra and A.~Melchiorri,
  arXiv:0901.0919 [astro-ph.CO].


\bibitem{Hu01c}
W. Hu,   Phys.\ Rev.\  D {\bf 65} (2002) 023003, \eprint{astro-ph/0108090}.

\bibitem{Kap03}
M. Kaplinghat, New Astronomy Review, 47, 893, 2003,\eprint{astro-ph/0307538}.

\bibitem{AcqBac05}
V. Acquaviva, C. Baccigalupi, 2005, \eprint{astro-ph/0507644}.

\bibitem{serra09}
  P.~Serra, A.~Cooray, S.~F.~Daniel, R.~Caldwell and A.~Melchiorri,
  arXiv:0901.0917 [astro-ph.CO].



\bibitem{KapKnoSon03}
M. Kaplinghat et al., Phys. Rev. Lett., 91, 241301,  \eprint{astro-ph/0303344}.


\bibitem{lensteo}
  M.~Zaldarriaga and U.~Seljak,
  Phys.\ Rev.\  D {\bf 58} (1998) 023003
  [arXiv:astro-ph/9803150];
  A.~Lewis and A.~Challinor,
  Phys.\ Rept.\  {\bf 429} (2006) 1
  [arXiv:astro-ph/0601594].

\bibitem{beandore}
  R.~Bean and O.~Dore,
  Phys.\ Rev.\  D {\bf 69} (2004) 083503
  [arXiv:astro-ph/0307100].


\bibitem{Lewis:2002ah}
A. Lewis and S. Bridle,
Phys.\ Rev.\ D {\bf 66}, 103511 (2002) (Available from
\texttt{http://cosmologist.info}.)

\bibitem{:2006uk}
    [Planck Collaboration],
  arXiv:astro-ph/0604069.

\bibitem{Baumann:2008aq}
  D.~Baumann {\it et al.},
  arXiv:astro-ph/0811.3919.

\bibitem{Ma:1995ey}
  C.~P.~Ma and E.~Bertschinger,
  Astrophys.\ J.\  {\bf 455}, 7 (1995).

\bibitem{Lewis:2005tp}
  A.~Lewis,
  Phys.\ Rev.\  D {\bf 71} (2005) 083008
  [arXiv:astro-ph/0502469].

\bibitem{Lewis:2006fu}
  A.~Lewis and A.~Challinor,
  Phys.\ Rept.\  {\bf 429}, 1 (2006).


\bibitem{alessandra}
  G.~B.~Zhao, L.~Pogosian, A.~Silvestri and J.~Zylberberg,
  arXiv:0905.1326 [astro-ph.CO].


\bibitem{HS}
  W.~Hu and I.~Sawicki,
  Phys.\ Rev.\  D {\bf 76}, 064004 (2007)
  [arXiv:0705.1158 [astro-ph]].

\bibitem{ermi08}
  E.~Calabrese, A.~Slosar, A.~Melchiorri, G.~F.~Smoot and O.~Zahn,
  Phys.\ Rev.\  D {\bf 77}, 123531 (2008)
  [arXiv:0803.2309 [astro-ph]].

\bibitem{shk06}
  K.~M.~Smith, W.~Hu and M.~Kaplinghat,
  Phys.\ Rev.\  D {\bf 74} (2006) 123002
  [arXiv:astro-ph/0607315];
C.~Li, T.~L.~Smith and A.~Cooray,
  Phys.\ Rev.\  D {\bf 75} (2007) 083501
  [arXiv:astro-ph/0607494].

\bibitem{shk04}
  K.~M.~Smith, W.~Hu and M.~Kaplinghat,
  Phys.\ Rev.\  D {\bf 70} (2004) 043002
  [arXiv:astro-ph/0402442].

\bibitem{huokamoto}
  T.~Okamoto and W.~Hu,
  Phys.\ Rev.\  D {\bf 67} (2003) 083002
  [arXiv:astro-ph/0301031];
  T.~Okamoto and W.~Hu,
  Phys.\ Rev.\  D {\bf 66} (2002) 063008
  [arXiv:astro-ph/0206155];
  A.~Cooray and M.~Kesden,
  New Astron.\  {\bf 8} (2003) 231
  [arXiv:astro-ph/0204068];
    M.~H.~Kesden, A.~Cooray and M.~Kamionkowski,
  Phys.\ Rev.\  D {\bf 67} (2003) 123507
  [arXiv:astro-ph/0302536].

\bibitem{hirataseljak}
  C.~M.~Hirata and U.~Seljak,
  Phys.\ Rev.\  D {\bf 68} (2003) 083002
  [arXiv:astro-ph/0306354].


\end{thebibliography}
\end{document}